\documentclass[twocolumn]{aastex61}

\received{May 3, 2019}
\revised{October 29, 2019}
\accepted{\today}
\submitjournal{ApJ}

\shorttitle{Evolution of an intermediate filament}
\shortauthors{Yardley et al.}

\begin{document}

\title{Understanding the plasma and magnetic field evolution of a filament using observations and nonlinear force-free field modelling}

\correspondingauthor{Stephanie L. Yardley}
\email{sly3@st-andrews.ac.uk}

\author[0000-0003-2802-4381]{Stephanie L. Yardley}
\affiliation{Mullard Space Science Laboratory, University College London, Holmbury St. Mary, Dorking, Surrey, RH5 6NT, UK}
\affiliation{School of Mathematics and Statistics, University of St Andrews, North Haugh, St Andrews, Fife, KY16 9SS, UK}

\author[0000-0002-5598-046X]{Antonia Savcheva}
\affiliation{Institute of Astronomy and National Astronomical Observatory, Bulgarian Academy of Sciences, 72 Tsarigradsko Chaussee Blvd., 1784 Sofia, Bulgaria}

\author[0000-0002-0053-4876]{Lucie M. Green}
\affiliation{Mullard Space Science Laboratory, University College London, Holmbury St. Mary, Dorking, Surrey, RH5 6NT, UK}

\author[0000-0002-2943-5978]{Lidia van Driel-Gesztelyi}
\affiliation{Mullard Space Science Laboratory, University College London, Holmbury St. Mary, Dorking, Surrey, RH5 6NT, UK}
\affiliation{Observatoire de Paris, LESIA, UMR 8109 (CNRS), F-92195 Meudon-Principal Cedex, France}
\affiliation{Konkoly Observatory of the Hungarian Academy of Sciences, Budapest, Hungary}

\author[0000-0003-3137-0277]{David Long}
\affiliation{Mullard Space Science Laboratory, University College London, Holmbury St. Mary, Dorking, Surrey, RH5 6NT, UK}

\author[0000-0001-9922-8117]{David R. Williams}
\affiliation{ESA European Space Astronomy Centre, 28692 Villanueva De La Ca\~{n}ada, Madrid, Spain.}

\author[0000-0001-6065-8531]{Duncan H. Mackay}
\affil{School of Mathematics and Statistics, University of St Andrews, North Haugh, St Andrews, Fife, KY16 9SS, UK}

\begin{abstract}

We present observations and magnetic field models of an intermediate filament present on the Sun in August 2012, associated with a polarity inversion line that extends from AR 11541 in the east into the quiet sun at its western end. A combination of {\it SDO}/AIA, {\it SDO}/HMI, and GONG H$\alpha$ data allow us to analyse the structure and evolution of the filament from 2012 August 4 23:00~UT to 2012 August 6 08:00~UT when the filament was in equilibrium. By applying the flux rope insertion method, nonlinear force-free field models of the filament are constructed using {\it SDO}/HMI line-of-sight magnetograms as the boundary condition at the two times given above. Guided by observed filament barbs, both modelled flux ropes are split into three sections each with a different value of axial flux to represent the non-uniform photospheric field distribution. The flux in the eastern section of the rope increases by 4$\times$10$^{20}$~Mx between the two models, which is in good agreement with the amount of flux cancelled along the internal PIL of AR 11541, calculated to be 3.2$\times$10$^{20}$~Mx. This suggests that flux cancellation builds flux into the filament's magnetic structure. Additionally, the number of field line dips increases between the two models in the locations where flux cancellation, the formation of new filament threads and growth of the filament is observed. This suggests that flux cancellation associated with magnetic reconnection forms concave-up magnetic field that lifts plasma into the filament. During this time, the free magnetic energy in the models increases by 0.2$\times$10$^{31}$~ergs.


\end{abstract}

\keywords{Sun: activity --- Sun: filaments, prominences --- Sun: coronal mass ejections (CMEs) --- Sun: evolution --- Sun: magnetic fields  --- Sun: photosphere}

\section{Introduction}

Filaments are accumulations of cool, dense, partially ionised plasma that are suspended in the solar corona against gravity. They lie above polarity inversion lines in the photospheric radial magnetic field (PIL; \citealt{Babcock-1955}). This includes the PIL of active regions (``active region filaments"), between active regions (``intermediate filaments") and in the quiet sun (``quiescent filaments").

When observed on disk in H$\alpha$, filaments are seen to have a main body that extends horizontally along the structure called a spine, and barbs, which are lateral extensions protruding from the spine at an acute angle (e.g. \citealt{Martin-1992, Martin-1994, Martin-1998, Lin-2008}). Both of these substructures exhibit thin threads of flowing plasma that are thought to outline the magnetic field \citep{Martin-2008}. The spine represents the axial magnetic field of the filament, whereas barbs extend vertically down to the chromosphere or possibly into the photosphere. Barbs can also be used to indirectly determine the chirality of a filament when viewed from the positive-polarity side, with left- (right-) bearing barbs being an indication that the filament has sinistral (dextral) chirality \citep{Martin-1994}.

Due to the high electrical conductivity of the corona, and even of the relatively weakly ionised prominence plasma, the plasma is frozen into the magnetic field, meaning that plasma can move freely along field lines but not across them. Under these conditions the magnetic field configuration is thought to play a major role in supporting filament plasma against gravity. This led \citet{Kippenhahn-1957} to suggest that dips in the magnetic field configuration of filaments could provide locations for plasma accumulation. Since then, filament models have evolved and are usually divided into two main groups. The first group involves a weakly-twisted magnetic field configuration, known as a flux rope \citep{Kuperus-1974, Pneuman-1983, vB-89}, where filament material is supported in concave-up sections of the magnetic field. The second group involves a sheared arcade in which field lines can have dips \citep{Antiochos-1994, Devore-2000}. However, it has been proposed that due to the dynamic nature of filament plasma, magnetic dips may not be necessary for their formation (e.g. \citealt{Martin-1994, Karpen-2001}).

In addition to understanding the specific magnetic configuration that can support filament mass, the physical processes that allow filament material to form and accumulate must also be addressed. There are a variety of physical mechanisms that could explain the accumulation of plasma. These include the emergence of a highly-twisted flux rope \citep{Rust-1994}, U-loop emergence \citep{Deng-2000}, or magnetic reconnection associated with the observation of flux cancellation that lifts plasma into the atmosphere \citep{vB-89,Litvinenko-1999, Litvinenko-2007, Litvinenko-2015}. Also, the direct injection of chromospheric plasma \citep{Poland-1986, Wang-1999,Chae-2001}, which is largely motivated by the connection between flux cancellation and the formation of filament channels \citep{vB-89,Martin-1998,Wang-2007,Wang-2013}, and evaporation-condensation models \citep{Engvold-1977, An-1985, Antiochos-1991, Dahlburg-1998}. For a more in-depth review of the magnetic structure and dynamics of filaments see \cite{Mackay-2010}.

The first 3D magnetic models of filaments using linear force-free field (LFFF) extrapolations of the photospheric line-of-sight (LoS) magnetic field were developed by \citet{Aulanier-1998a}. These magnetic models were able to explain many observed features of H$\alpha$ filaments, in particular, the orientation and hence chirality patterns of the filament barbs and also the vector magnetic field measurements. In the model the barbs are formed by concave-up dips in the magnetic field that are local to small-scale PILs of parasitic (minority) polarities in the photosphere. These sites correspond to the existence of field that is tangential to the photosphere known as ``bald patches" \citep{Titov-1993}. Furthermore, \citet{Aulanier-1998b} and \citet{Mackay-2009} found that the motion of the filament barbs corresponds to the changes of the parasitic polarities. This 3D modelling of filaments supports the interpretation that the plasma is supported in the magnetic field configuration of a flux rope.

The comparison of observed photospheric magnetic field under filaments with coronal modelling, indicates that flux cancellation might be fundamentally important to the formation and evolution of filaments. Flux cancellation is seen as small-scale opposite polarity features that converge, collide, and subsequently disappear along the PIL \citep{Martin-1985}. The opposite-polarity features that collide and subsequently disappear are interpreted as representing the footpoints of two magnetic flux systems, which become sheared across the PIL. During collision, these features are assumed to undergo magnetic reconnection that leads to the formation of two magnetic flux systems with a different connectivity to the pre-reconnection pair: a small loop which is pulled below the photosphere by magnetic tension (the observational manifestation of flux cancellation) and a longer, highly sheared loop that remains in the solar atmosphere as the flux rope axis. Once a sufficient amount of flux has accumulated at the PIL this process starts to build helical field around the axis, forming the flux rope. Therefore, ongoing flux cancellation associated with the process of magnetic reconnection can transform a sheared coronal field into a flux rope \citep{vB-89}.
 
 
The amount of flux available to be built into the rope is equal to the total amount of flux cancelled. However, the amount of flux that is built into the rope may differ from the amount of flux cancelled, depending upon properties such as the amount of shear and length of the PIL along which flux cancellation is taking place \citep{Green-2011}. The concave-up sections of the flux rope that move through the dense plasma of the lower atmosphere not only provide locations that are capable of supporting dense filament plasma but also allow plasma to be pulled into the rope.

Nonlinear force-free field (NLFFF) models are more suitable than LFFF to describe the configuration of the non-potential magnetic field of a flux rope, which is held down by an overlying potential arcade. Static NLFFF models created at different times during the evolution of a region can provide its 3D magnetic field structure, as field lines from the models can be verified using observed plasma emission and absorption structures. This allows us to investigate, for example, whether a modelled flux rope, and its corresponding features, are consistent with the observations. To date, the evolution of several active regions that exhibit filaments have been modelled applying the magnetofrictional relaxation technique \citep{Yang-1986}. This technique uses the photospheric LoS magnetic field as the boundary condition for the extrapolation, observations of a filament to guide the position of an inserted flux rope and EUV/X-ray emitting loops to constrain the model. (e.g. \citealt{Bobra-2008,Su-2009,Savcheva-2012}). This method also allows certain topological features of the magnetic field to be studied such as the presence of magnetic dips or bald patches, which can be compared with photospheric magnetic field and coronal observations. It is also possible to use the magnetofrictional relaxation technique to construct a continuous time-dependent series of NLFFF models by evolving the initial coronal magnetic field through changing the lower boundary conditions \citep{Mackay-2011, Gibb-2014, Yardley-2018a, Yardley-2019}. Another set of studies invoke a more general method to create static models of flux ropes in active regions by using the photospheric vector magnetic field to constrain the NLFFF models (e.g. \citealt{Regnier-2002,Schrijver-2006, Canou-2010, Guo-2016}). The signatures of magnetic dips have also been observed in vector magnetic field data as bald patches, where the magnetic field is tangent to the photosphere and crosses the PIL in the inverse direction \citep{Lites-2005, Lopez-2006, Okamoto-2008, Yardley-2016}. That said, both these methods are affected by uncertainties in the direction of the transverse field due to the 180$^{\circ}$ ambiguity and so bald patches must be spatially and temporally coherent.

These previous NLFFF studies have all focused on modelling active region filaments, which are reasonably compact and located along PILs in a strong magnetic field distribution. In active regions the transverse component of the field normally exceeds the level of noise associated with this field component. This is not necessarily the case for observations of the transverse component in the quiet sun. Therefore, to study the magnetic structure of filaments outside active regions NLFFF models that require only the LoS photospheric magnetic field, due to its lower noise values, can be used. For example, \citet{Su-2012} used the flux rope insertion method to model a polar crown filament as this method relies only on the LoS magnetic field component. In a more recent study \citet{Jiang-2014} utilized the vector magnetic field extrapolation technique to model the coronal flux rope of a large-scale intermediate filament located between an active region and a weak field region. The study successfully managed to match the magnetic dips in a flux rope to  observations of the filament and its barbs.

In this paper we present the study of an intermediate filament that was observed on the Sun during August 2012. We model the magnetic field of the filament at two different times using the flux rope insertion method \citep{vB-2004} and compare these models to photospheric, chromospheric and coronal observations. This is the first time the flux rope insertion method has been used to model a filament of this nature. The intermediate filament has its eastern end anchored along the internal PIL of AR 11451, which is situated in an active region complex (ARs 11538--11541) and the western end positioned in the quiet sun. In order to further our understanding of the physical processes and magnetic field configuration responsible for the formation and support of filament plasma a comparison is made between the observations and NLFFF models over a 33 hour time period. A NLFFF model is constructed at both the start and end of this time period, allowing us to investigate the locations of dips in the modelled field, the flux content of the magnetic structure of the filament and the variation of these properties with time. These are compared with the sites of flux cancellation and the evolution of the filament plasma distribution.

The paper is structured as follows. Section~\ref{sec:ins} gives details on the instrumentation used and the algorithm application for observational analysis of flux cancellation. Section~\ref{sec:obs} describes the photospheric field distribution and chromospheric and coronal observations of the intermediate filament. In Section~\ref{sec:non} details of the flux rope insertion method, used to construct the two NLFFF models, are given. Section~\ref{sec:res} provides the results and the comparison of the observations with the NLFFF models. Finally, the results are discussed and the conclusions are given in Section~\ref{sec:sum}.

\section{Instrumentation \& Algorithm Application} \label{sec:ins}

\subsection{Instrumentation}

The evolution of the intermediate filament and the associated photospheric magnetic field are analyzed in detail during the period 2012 August 4--6 using a wide range of space-borne and ground-based instrumentation. A brief description of the data used is now given.

Data taken by the Extreme UltraViolet Imager on board the Solar TErrestrial RElations Observatory-B (STEREO-B/EUVI; \citealt{Wuelser-2004, Howard-2008}) are used to calculate the height of the filament when it is seen at the west limb from the STEREO-B viewpoint. At this time the filament is near Sun-centre from the AIA perspective. The evolution and dynamics of the filament plasma are studied during its disk passage using the Atmospheric Imaging Assembly (AIA; \citealt{Lemen-2012}) on board the \textit{Solar Dynamics Observatory} ({\it SDO}; \citealt{Pesnell-2012}). AIA provides observations of multi-temperature plasma, which allows us to investigate the plasma evolution in the chromosphere and corona with respect to changes in the photospheric magnetic field. In particular, we focus on observing the plasma evolution in the 171, 193, 211 and 304~{{\AA}} wavebands, which are dominated by plasma temperatures around 0.6~MK, a combination of 1.2 and 20~MK, 2~MK and 0.05~MK, respectively. H$\alpha$ images taken from the Global Oscillation Network Group (GONG; \citealt{Harvey-2011}) are also used to study the filament evolution and the orientation of its barbs to determine the chirality sign of the filament. The evolution of the photospheric magnetic field is analyzed using full disk LoS magnetogram data from the Helioseismic and Magnetic Imager (HMI; \citealt{Schou-2012}) on board {\it SDO}. In this study, we focus on analyzing observations of the filament during the time period beginning 2012 August 4 at 23:00~UT until 2012 August 6 at 08:00~UT when the filament remains  in equilibrium. However, to constrain the magnetic field models we use observations taken when the filament is activated as during these times heated plasma reveals important information about the magnetic field configuration. The occurrence and timings of the CMEs associated with the filament are determined from observations made by the Large Angle and Spectrometric Coronagraph (LASCO, \citealt{Brueckner-1995}) on board the \textit{Solar and Heliospheric Observatory} ({\it SOHO}; \citealt{Domingo-1995}). Filament activations and eruptions have been previously studied by \citet{Li-2013, Srivastava-2013, Joshi-2014}.

\subsection{Algorithm Application} \label{sec:AA}

The photospheric field evolution of AR 11541 and the quiet sun photosphere above which the filament resides is studied using the HMI 720~s cadence data series \citep{Hoeksema-2014, Couvidat-2016}.  The flux cancellation is quantified by using the Solar Tracking of the Evolution of photospheric Flux (STEF) algorithm, the discussion of which can be found in \citet{Yardley-2016, Yardley-2018b}. Once the region for analysis is identified the radial component of the magnetic field is estimated by applying a cosine correction using the Heliocentric Earth Equatorial (HEEQ) coordinate system \citep{Thompson-2006}. The radialised field data are then differentially rotated to the central meridian passage time using a routine that corrects for area foreshortening. For this particular case no smoothing is applied and pixels are selected with magnetic flux density values above a threshold of 3$\sigma$, where $\sigma$ is the noise in the LoS magnetic field, which is 10~G for HMI. No smoothing and a low threshold is applied to the data in this case as the field of the decayed active region and the quiet sun is quite dispersed.
Due to the dispersed nature of the photospheric magnetic field along the filament, even in the active region section, a second criterion is applied.
That is, magnetic fragments must be 4 HMI pixels or larger to be selected for measurement to avoid false detections. Once this process is completed the total magnetic flux is calculated from the selected pixels. Only flux cancellation occurring in AR 11541 is quantified, by calculating the reduction in the total positive magnetic flux in the active region since the negative magnetic flux is more dispersed and difficult to measure. Flux cancellation sites are identified along the section of the filament that extends into the quiet sun, but the difficulty of linking a flux cancellation event with the magnetic structure of the filament precludes a quantitative analysis. 
To separate out the different flux cancellation sites for both the ARs and quiet sun a mask was applied to the magnetograms.

\section{Observations} \label{sec:obs}

\subsection{Photospheric Field \& Filament Evolution} \label{obs}

The filament initially forms during Carrington Rotation (CR) 2125,
remains intact until it rotates back onto the disk during CR 2126 on 2012 August 1, and makes its final appearance during CR 2127 when it finally erupts on 2012 August 31. During its lifetime there are four eruptions: two eruptions that lead to the destabilization of the filament and two eruptions which result in the ejection of filament material. In this study we concentrate on analyzing observational data taken during CR 2126, in particular, a time period after one of the eruptions when the filament is stable beginning on 2012 August 4 at 23:00~UT until 2012 August 6 at 08:00~UT, after which the filament gets destabilized and erupts again. This enables us to study the filament evolution and associated flux cancellation during a quiet period.

\begin{figure}
\plotone{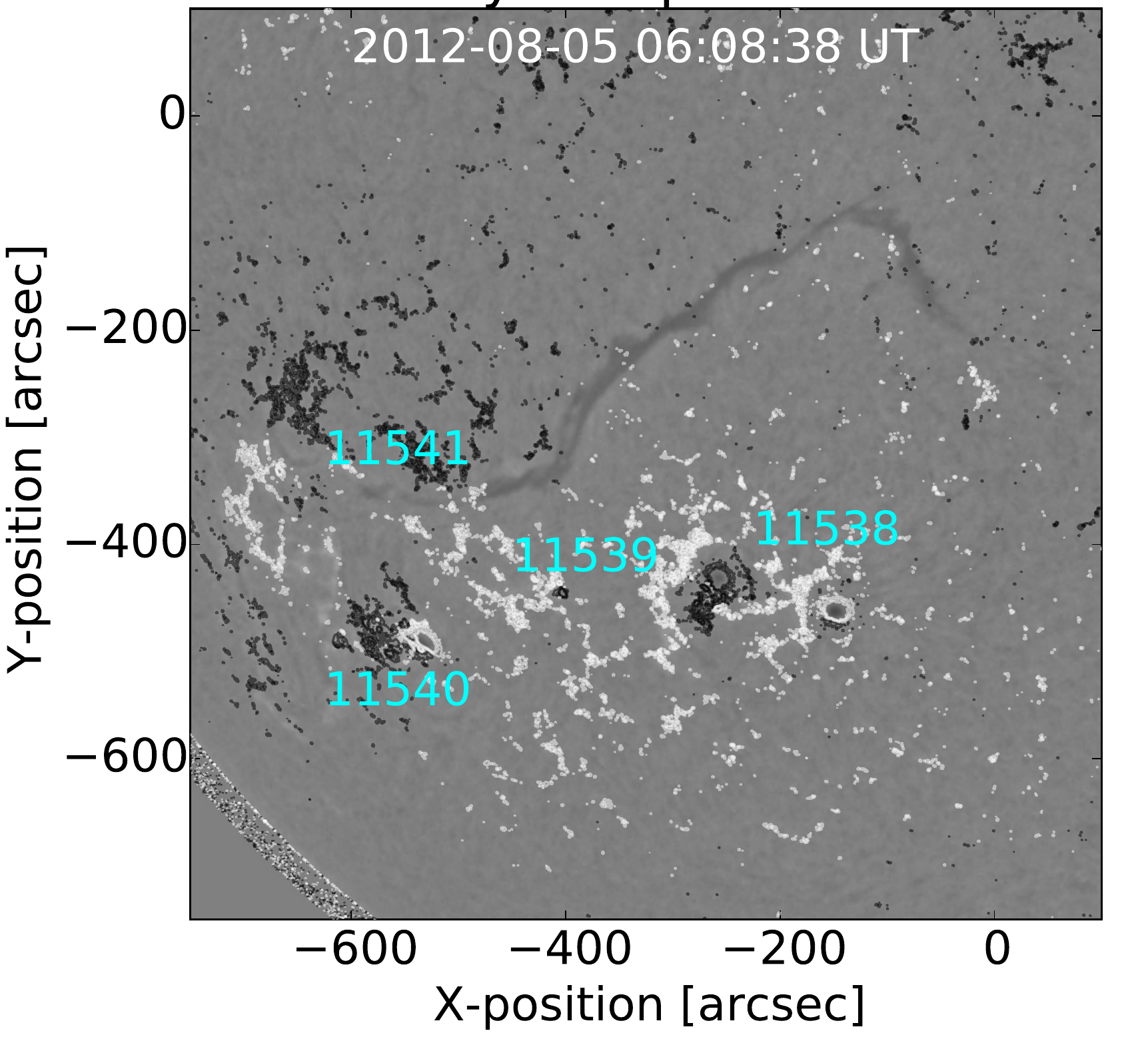}
\caption{H$\alpha$ image from the GONG Network shows the filament and locations of the four ARs in the AR complex (ARs 11538--11541), which have been labelled in blue. The corresponding \textit{SDO}/HMI magnetogram has been overlaid on the image where white (black) contours represent the positive (negative) magnetic field with saturation levels of $\pm$100~G. \label{fig1}
}
\end{figure}

The eastern end of the filament is rooted in an AR complex that includes decayed regions (AR 11519, 11520 and 11521) from the previous rotation (CR 2125) and a new region that emerged on the far-side. These four ARs are numbered: 11538, 11539, 11540 and 11541 (in CR 2126). The filament exists along a PIL that begins in the western section of AR 11541 and extends into the quiet sun (Figure~\ref{fig1}).

\begin{figure*}[t]
\epsscale{1}
\plotone{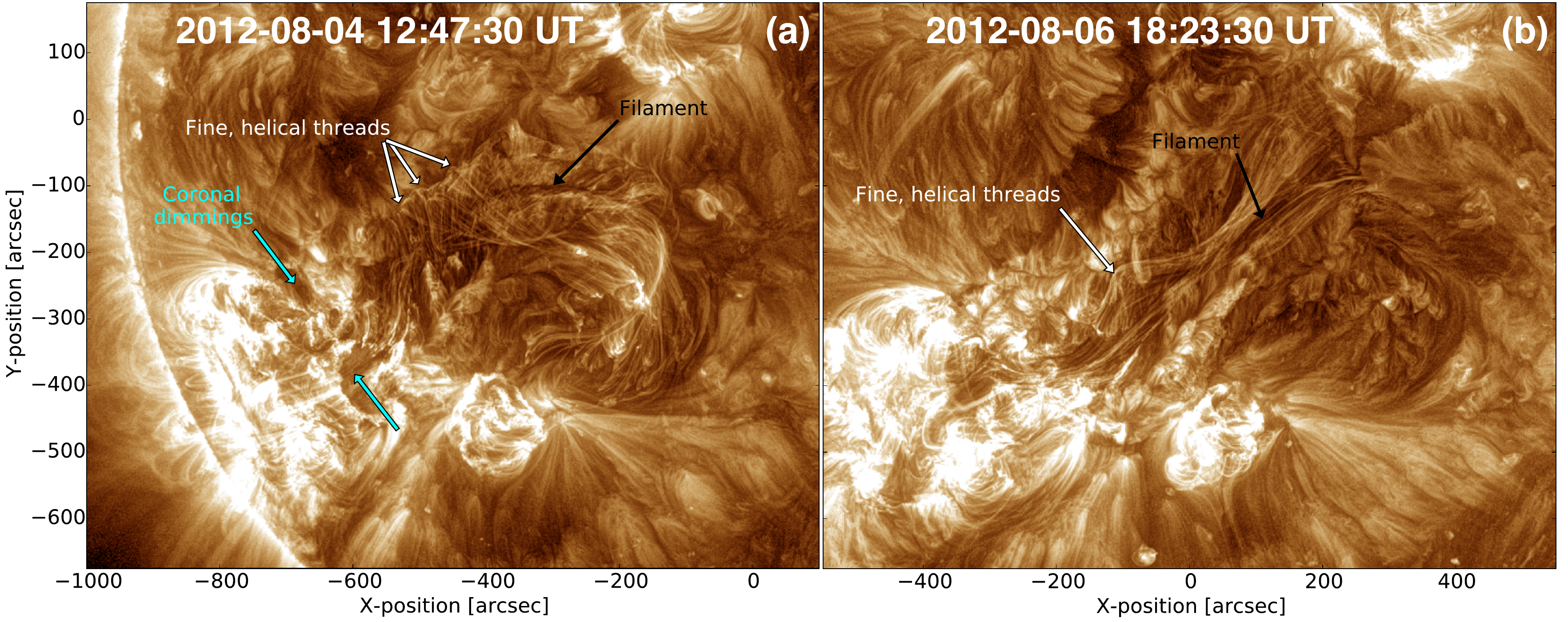}
\caption{\textit{SDO}/AIA 193~{\AA} images of the filament that have been processed using the Multi-scale Gaussian Normalisation (MGN) technique of \citet{Morgan-2014}. Panel (a): The filament (indicated by the black arrow) is perturbed by the erupting structure in the south east. The corresponding coronal dimmings have been labelled with blue arrows. Panel (b): The filament becomes activated for a second time. The white arrows in both panels indicate the fine, helical threads that show the filament is supported by a twisted structure. An animation of this figure is available online. \label{fig2}}
\end{figure*}

The two eruptions that act to destabilize the filament are described below as the EUV observations taken during these times (see Figure~\ref{fig2}) are used to constrain the NLFFF models of the filament.


On 2012 August 4 at approximately 11:12~UT the filament becomes activated in response to the eruption of a structure overlying the filament's eastern section in AR 11541. Double coronal dimmings, which are an indication of the footpoints of the erupting magnetic configuration, are observed to be situated over the magnetic polarities of AR 11541, either side of the flare arcade (Figure~\ref{fig2} (a)). The interaction between the erupting structure and the filament leads to the perturbation and heating of the filament, revealing plasma that follows helical magnetic field lines. The presence of fine helical plasma threads suggests that the filament plasma is supported in a flux rope configuration. The right-handed twist of the plasma threads is consistent with a flux rope of positive chirality. The eruption, which is associated with a GOES C3.5 class flare, is observed by SOHO/LASCO C2 at around 12:48~UT and also by STEREO-B/SECCHI COR1 (13:35~UT) and COR2 (14:24~UT). A more detailed study of the eruption and its impact on the filament is presented by \citet{Joshi-2014}. The filament is stable again by 17:00~UT on August 4.
On August 6 at approximately 08:30~UT the middle section of the filament activates and the western end begins to rise around 12:45~UT. Fine threads of flux rope plasma become heated during this period revealing helical threads in the central section of the filament (Figure~\ref{fig2} (b)). A fraction of the filament material appears to lift off slowly at the western end from around 14:00~UT. There is a C1.1 GOES class flare at 19:50~UT in AR 11541, which is associated with a faint, slow CME visible in LASCO/C2 at 20:24 UT.

The majority of the filament barbs, identified in the H$\alpha$ data, are left-bearing indicating that the filament has sinistral chirality, which is typical for the southern hemisphere and is in agreement with \citet{Joshi-2014}.

The height of the filament plasma is estimated using STEREO-B, which is positioned at 114.8$^{\circ}$ away from the Sun-Earth line, trailing the Earth. The central section of the filament is seen to be suspended directly above the west limb from the STEREO-B perspective. The height of the plasma is measured at the time of the first NLFFF model (2012 August 4 at 23:00~UT). The filament plasma spans a range of heights from approximately 7~Mm to 47~Mm above the photosphere.

\begin{figure*}[t]
\epsscale{1}
\plotone{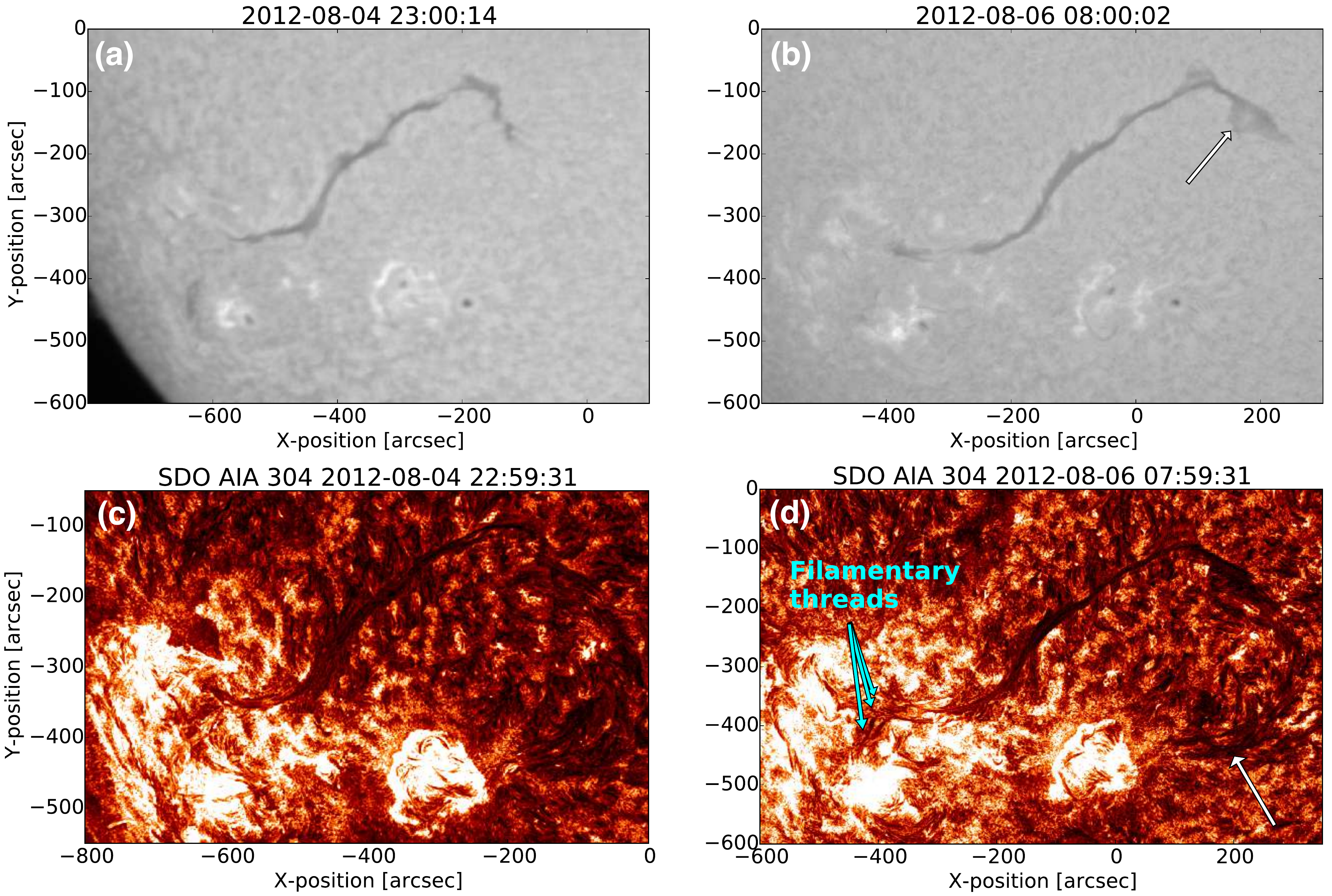}
\caption{Top row: H$\alpha$ images of the filament taken by the GONG network on 2012 August 4 at 23:00~UT (panel (a)) and 2012 August 6 (panel (b)) corresponding to the times of the NLFFF models. The growth of the western end of the filament is indicated by the white arrow in panel (b). Bottom row: 304~${\AA}$ images of the filament that have been enhanced using the MGN technique of \citet{Morgan-2014}. (Panel (c) shows the filament on 2012 August 4 at 23:00~UT at the time of the first NLFFF model. Panel (d) shows an image of the filament taken on 2012 August 6 at 08:00~UT at the time of the second NLFFF model. The blue and white arrows indicate the formation of new filamentary threads and the extension of the western end of the filament towards AR 11538. \label{fig3}}
\end{figure*}

The filament is seen to exhibit plasma flows both along and perpendicular to its spine. These flows are best observed in the 304~{{\AA}} waveband of AIA and are concentrated in the eastern section of the filament that overlies the PIL of the active region. These include counter-streaming flows that are most apparent on August 5 and 6. The filament is also observed to have grown in size between August 4 23:00~UT and August 6 08:00~UT. In particular, the extension of the filament's western end towards AR 11538 is notable (see white arrow in Figure~\ref{fig3} (b) and the blue arrows in Figure~\ref{fig3} (d)). There are also more filamentary threads present (blue arrows in Figure~\ref{fig3} (d)) along and perpendicular to the internal PIL of AR 11541.

\subsection{Flux Cancellation}

\begin{figure*}[t]
\epsscale{1}
\plotone{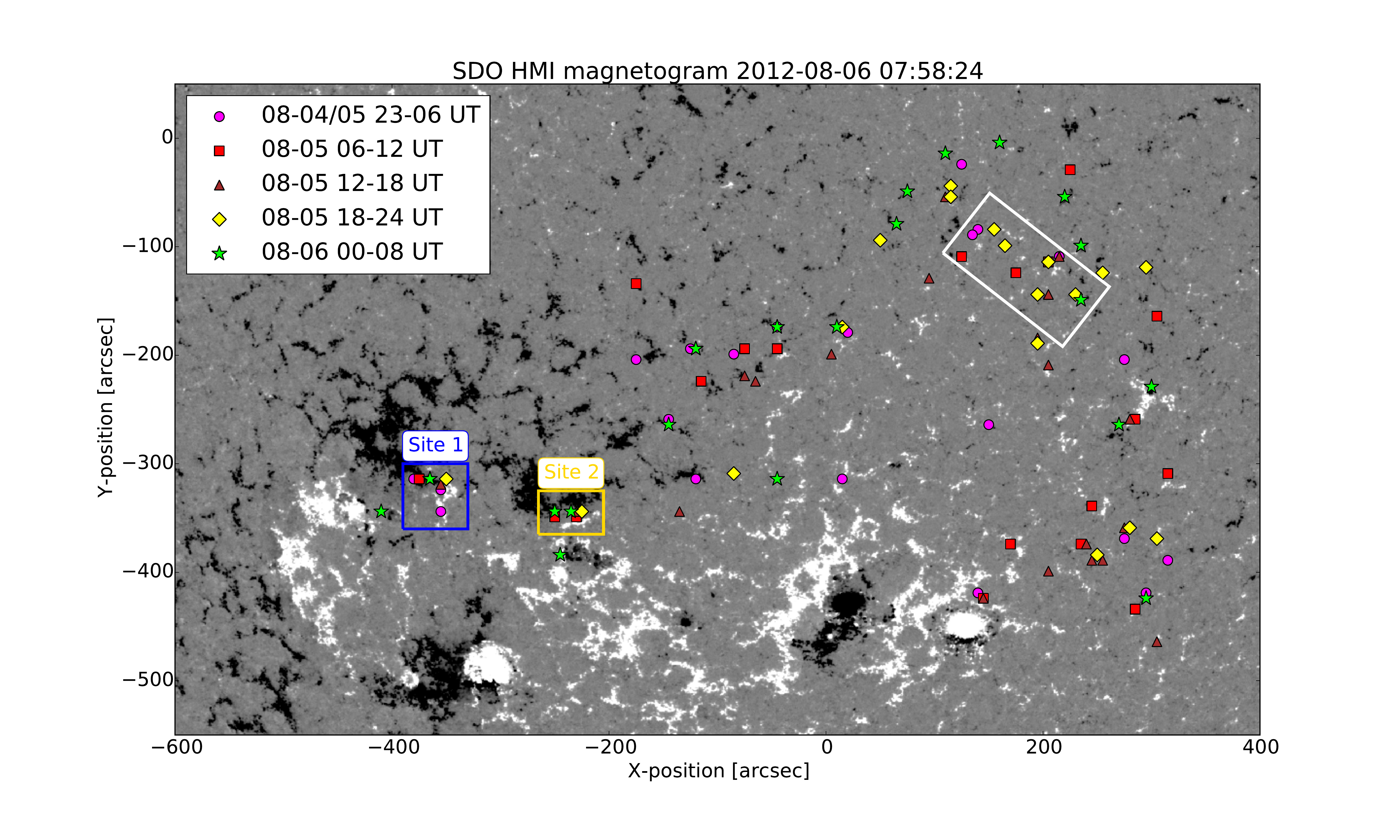}
\caption{The location of flux cancellation sites plotted on the \textit{SDO}/HMI magnetogram taken on 2012 August 6 at 07:58~UT. The magnetogram has been differentially rotated to 2012 August 6 at 17:00~UT when the middle section of the filament crosses central meridian. The different markers represent cancellation occurring during various 6~hour time periods given in the legend. The two sites (1 and 2), located along the internal PIL of AR 11541, labelled with blue and yellow boxes respectively, are the sites where flux cancellation is quantified. The white box indicates the cancellation sites that are observed underneath the section of the filament rooted in the quiet sun. \label{fig4}}
\end{figure*}

The photospheric magnetic field data show that there are flux cancellation sites along the full length of the filament throughout the time period studied (2012 August 4 23:00~UT -- 2012 August 6 08:00~UT); both along the internal PIL of AR 11541 and in the quiet sun. There are also flux cancellation events in the immediate surrounding area of the filament channel. The flux cancellation sites are shown in Figure~\ref{fig4}; these sites are more abundant in the quiet sun than in the AR due to the spatial extent of the filament.


In the quiet sun, flux emergence and cancellation were observed to occur every few hours and there was no overall trend in the evolution of the magnetic field. Flux cancellation in the quiet sun has not been quantified in this study due to the difficulties in identifying which cancellation events are connected to the filament's magnetic structure. However, there were a total of 12 cancellation sites spanning an area of 5000~Mm$^{2}$ underneath the western section of the filament (white box in Figure~\ref{fig4}). The location of these cancellation sites corresponds to a section of the filament that grows in size as seen in the H$\alpha$ and AIA observations (white arrow in top-right panel of Figure~\ref{fig3}).

There are two main regions of ongoing flux cancellation located at the internal PIL of AR 11541. These two regions are indicated by blue and yellow boxes in Figure~\ref{fig4} and are referred to as sites 1 and 2, respectively. To calculate the quantity of flux cancelled in these two locations over the time period studied (2012 August 4 23:00~UT until 2012 August 6 08:00~UT) the STEF algorithm was applied. The quiet sun method was used due to the very dispersed and fragmented nature of the active region field (see Section~\ref{sec:AA}). It was not possible to isolate and determine the boundary of the negative polarity in this case so the amount of flux cancelled was calculated using the reduction in the total positive magnetic flux only. The total positive flux cancelled was calculated separately in sites 1 and 2. In addition to flux cancellation, a flux emergence episode occurred on August 5 between 14:00~UT and 15:48~UT at site 1 and on August 6 between 05:00~UT and 08:00~UT at site 2. These flux emergence events act to mask the flux cancellation occurring along the internal PIL of AR 11541. Therefore, the positive magnetic flux of the two emergence episodes are subtracted from the total positive flux. The amount of flux cancellation that occurred at sites 1 and 2 during the time period studied is 1.4 and 1.8~$\times$~10$^{20}$~Mx, respectively. Therefore, the total flux cancelled is 3.2~$\times$~10$^{20}$~Mx. This gives an average flux cancellation rate of 9.5~$\times$~10$^{18}$~Mx~h$^{-1}$.

\section{The NLFFF Model} \label{sec:non}

To study the 3D magnetic structure that supports the filament plasma we require the construction of coronal magnetic field models. This is due to the difficulty of directly measuring the coronal magnetic field and the presence of projection effects due to the contributions of multiple plasma structures along the line of sight in an optically thin plasma.

In this section, we discuss the construction of 3D NLFFF models of the intermediate filament at the start and end of our time period of study. We compare the magnetic field configuration of the models to the \textit{SDO}/AIA and HMI observations to determine the two ``best-fit" models; one on 2012 August 4 at 23:00~UT and the other on 2012 August 6 at 08:00~UT. These models are then used for subsequent analysis. In particular, the models are used to investigate the magnetic configuration of the filament. The models are also used to determine whether the observed flux cancellation sites correspond to the location of dips in the modelled field, and whether their evolution from the first model to the second is as expected considering the flux cancellation scenario. 

\subsection{Flux Rope Insertion Method}

Filaments exist in sheared and possibly twisted field configurations that are constrained by the overlying coronal magnetic field and as such are best described by NLFFF models. The magnetic field $\bf{B}$, of this configuration is given by the force-free criterion that can be derived from Amp{\'e}re's Law:  $\nabla \times \bf{B} \approx \alpha(\bf{r}) \bf{B}(\bf{r})$, where the torsion parameter $\alpha$, is constant along field lines but can vary as a function of position (i.e. field lines can have different values of $\alpha$). To model the non-potential coronal magnetic field we use the flux rope insertion method \citep{vB-2004}. Magnetic models are constructed by inserting a weakly-twisted flux rope (typically 1--1.5 turns) into a potential field extrapolation of the region. The potential field is constructed by applying the potential force source surface (PFSS) model using line-of-sight magnetograms taken by \textit{SDO}/HMI as the lower boundary condition. For the computational domain where the flux rope is inserted we use a high-resolution HMI magnetogram taken at the time of the two models (2012 August 4 at 23:00~UT and 2012 August 6 at 08:00~UT). We use a synoptic magnetogram from Carrington Rotation (CR) 2126 to construct the low-resolution global potential field that provides the side boundary conditions for the high-resolution domain. The method uses a staggered grid to ensure second order accuracy and to satisfy the solenoidal condition. The flux rope insertion method has been previously described in detail in papers such as \cite{Savcheva-2012} and \cite{Su-2012}, where it has been used to model sigmoids, active region and quiescent filaments. This is the first time the method has been used to model an intermediate filament.

The potential magnetic field model is modified by creating a cavity above and along the filament's path as determined from the 304~{\AA} AIA observations of filament plasma. The axial field of a flux rope is inserted into this cavity. Poloidal field is introduced by inserting a set of closed field lines that wrap around the axial field. This magnetic field configuration is not in equilibrium and so needs to be relaxed in order to reach a force-free equilibrium. This is achieved by applying magnetofrictional relaxation \citep{Yang-1986} along with hyperdiffusion \citep{Boozer-1986, Bhattacharjee-1986}. During the initial stages of relaxation resistive diffusion is used to merge the axial and poloidal fields of the flux rope together. The coronal field is then evolved using the induction equation, and the flux rope to expands until the magnetic tension of the overlying arcade balances the magnetic pressure associated with the flux rope. A small amount of hyperdiffusion is used to smooth out gradients in the force-free parameter while conserving magnetic helicity. This is an iterative process and we perform at least 30,000 iterations for each flux rope model.

\section{Model Results \& Comparison with Observations} \label{sec:res}

\begin{figure*}[t]
\epsscale{1}
\plotone{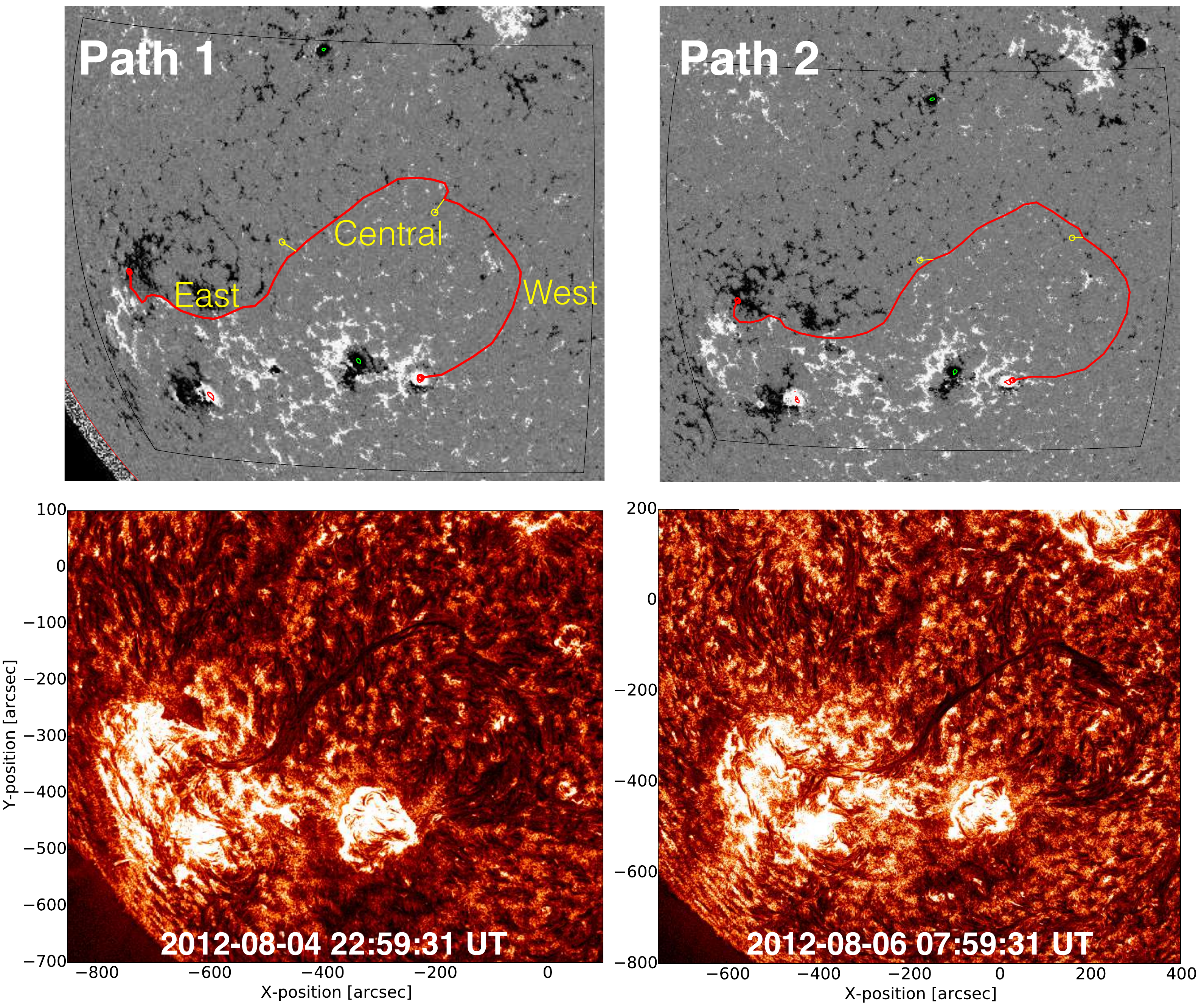}
\caption{\textit{SDO}/HMI magnetograms and the corresponding \textit{SDO}/AIA 304~{\AA} images of the filament taken around the times of the two magnetic models (2012 August 4 23:00~UT and August 6 08:00~UT). The magnetograms have a saturation of $\pm$100~G and the 304~{\AA} images have been enhanced using the MGN technique. The red curves represent the path along which the flux rope is inserted into the model. The path is divided into three sections: east, central and west as determined by the locations of the two main barbs of the filament (yellow bars in the top panel). The first section is at the eastern end where the flux rope is rooted in the negative polarity of AR 11541, the middle section is between the two barbs and the final section is between the second barb and the end rooted in the positive polarity of AR 11538. Each section has a different value of axial flux, which is given in Table~\ref{table:tab1}. }
\label{fig5}
\end{figure*}

We computed 26 models for each of the times 2012 August 4 at 23:00~UT and 2012 August 6 at 08:00~UT. The models have been constructed by inserting flux ropes with different combinations of of axial and poloidal fluxes in the ranges [0.5~$\times$10$^{20}$, 1~$\times$~10$^{21}$]~Mx and [5~$\times$~10$^{8}$, 1~$\times$10$^{10}$]~Mx~cm$^{-1}$, respectively. Figure~\ref{fig5} shows the filament paths for the first and second set of flux rope models.
The inserted flux rope has right-handed twist in accordance with the sinistral chirality of the filament, as the majority of the filament barbs are seen to be left-bearing in the observations (see Figure~\ref{fig1} and Section~\ref{obs}). The eastern footpoint of the filament is rooted in the negative polarity of AR 11541 whereas, the western end resides in the positive polarity of AR 11538. Therefore, the axial field of the flux rope points to the east. The eastern section of the filament is located in a region of stronger photospheric magnetic field compared to the western section therefore, the flux along the filament is likely to be non-uniform. This is reflected in the construction of the NLFFF models, which involves three sections each containing a different value of axial flux in order to keep the filament in equilibrium. These sections are constrained by the two main filament barbs that are represented by the yellow bars in Figure~\ref{fig5}. These sections are referred to as the eastern, central and western sections.

\subsection{Best-fit Models}

Once the NLFFF models are constructed, AIA observations are used to select the ``best-fit" model at each model time. Sample magnetic field lines from each model are selected and compared with AIA observations of the filament. The sample field lines were compared to the fine helical plasma threads, which are visible in 193~{\AA} during the time when the filament becomes activated (on August 4 between $\sim$12:00 and 15:00~UT, see Figure~\ref{fig2} and Section~\ref{obs}). In addition, plasma flows observed at the eastern end of the filament in 304~{\AA} were also used to identify aspects of the magnetic field configuration and help constrain the models. Finally, the chosen models had to possess coherent magnetic dips that overlay the axis of the flux rope. We note that the AIA 193~{{\AA}} observations used to select the best-fit models are taken during a time period when the filament is not in equilibrium. This approach was necessary as there were no overlying loops visible in the X-ray or EUV data that could be used to constrain the models during the phase where the filament is in equilibrium. The models are produced roughly 12~hours following and 6~hours preceding filament activation events caused by eruptions occurring in AR 11541.

\begin{table*}[ht]
	\vspace{10pt}
	\begin{tabular}{c c c c c}
		\hline
		\hline
		Model Times & & Axial Flux (10$^{20}$~Mx) & & Poloidal Flux (10$^{8}$~Mx~cm$^{-1}$)\\
		(UT) & East & Central & West \\
		\hline
	    04.08.12 23:00 & 1 & 1.5 & 0.5 & 5 \\
	    06.08.12 08:00 & 5 & 3 & 1 & 5 \\
		\hline
	\end{tabular}
	\caption{The axial and poloidal flux values initially inserted to construct the best-fit models on 2012 August 4 at 23:00~UT and on 2012 August 6 at 08:00~UT. The different axial flux values are given for the three sections of the filament as described in the text along with the values of poloidal flux. } \label{table:tab1}
\end{table*}

Table~\ref{table:tab1} gives the approximate values of axial flux of the three different sections of the flux rope (from east to west) for the two best-fit models. 
Both of the best-fit models have a poloidal flux of 5~$\times$10$^{8}$~Mx~cm$^{-1}$.
The axial flux in the eastern section of the flux rope increases by 4~$\times$~10$^{20}$~Mx from the time of the first NLFFF model to the time of the second model. This is comparable to the total amount of flux cancelled, as determined from observations, of 3.2~$\times$~10$^{20}$~Mx. Therefore, we find consistency between the NLFFF models and the photospheric evolution. free magnetic energy also increases from 8.8 $\times$10$^{31}$~ergs in model 1 to 1$\times$10$^{32}$~ergs in model 2, which is sufficient to power the eruption that occurs roughly 12 hours after model 2.

\begin{figure*}[t]
\epsscale{1}
\plotone{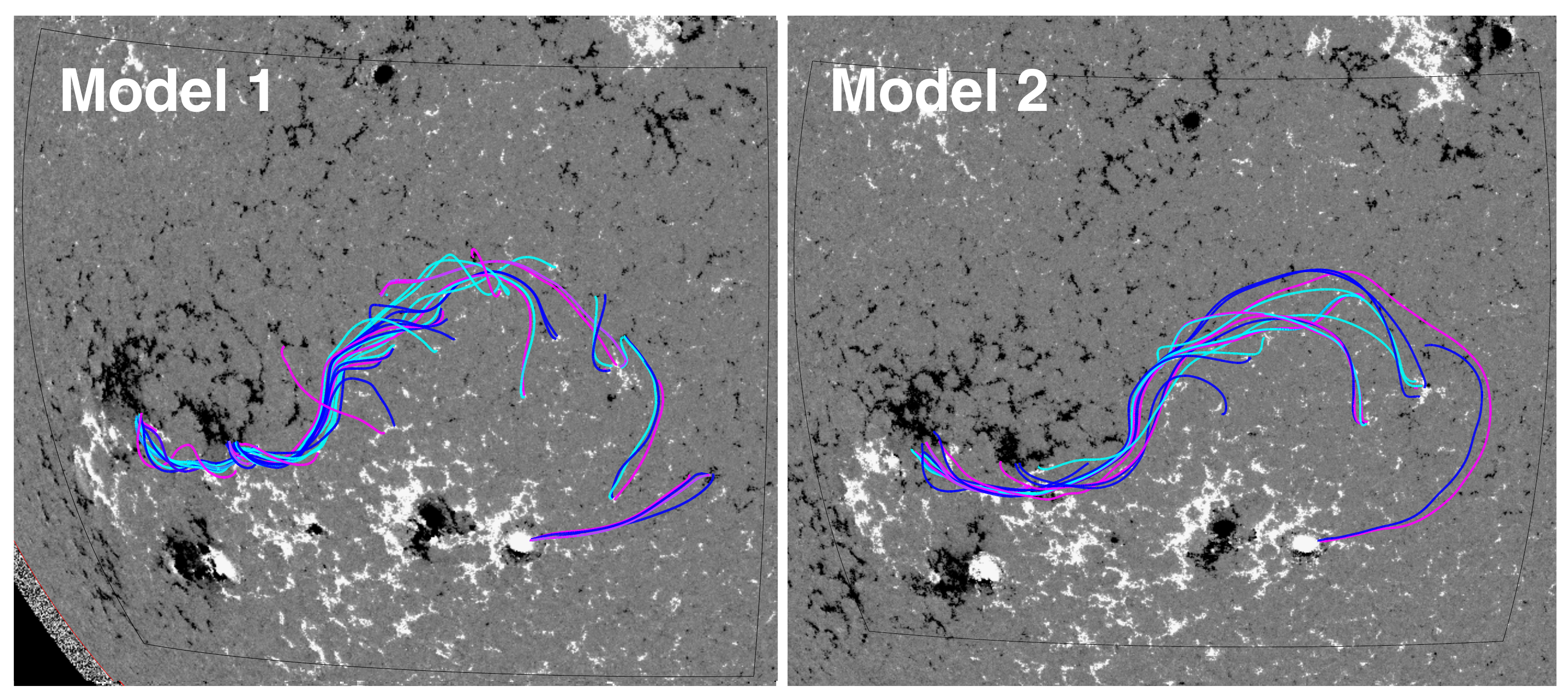}
\caption{Sample field lines taken from the NLFFF models constructed on 2012 August 4 at 23:00~UT (model 1) and 2012 August 6 08:00~UT (model 2). The field lines have been overlaid on the corresponding \textit{SDO}/HMI magnetograms, which are displayed with saturation levels of $\pm$100~G. } \label{fig6}
\end{figure*}

\begin{figure*}[t]
\epsscale{1}
\plotone{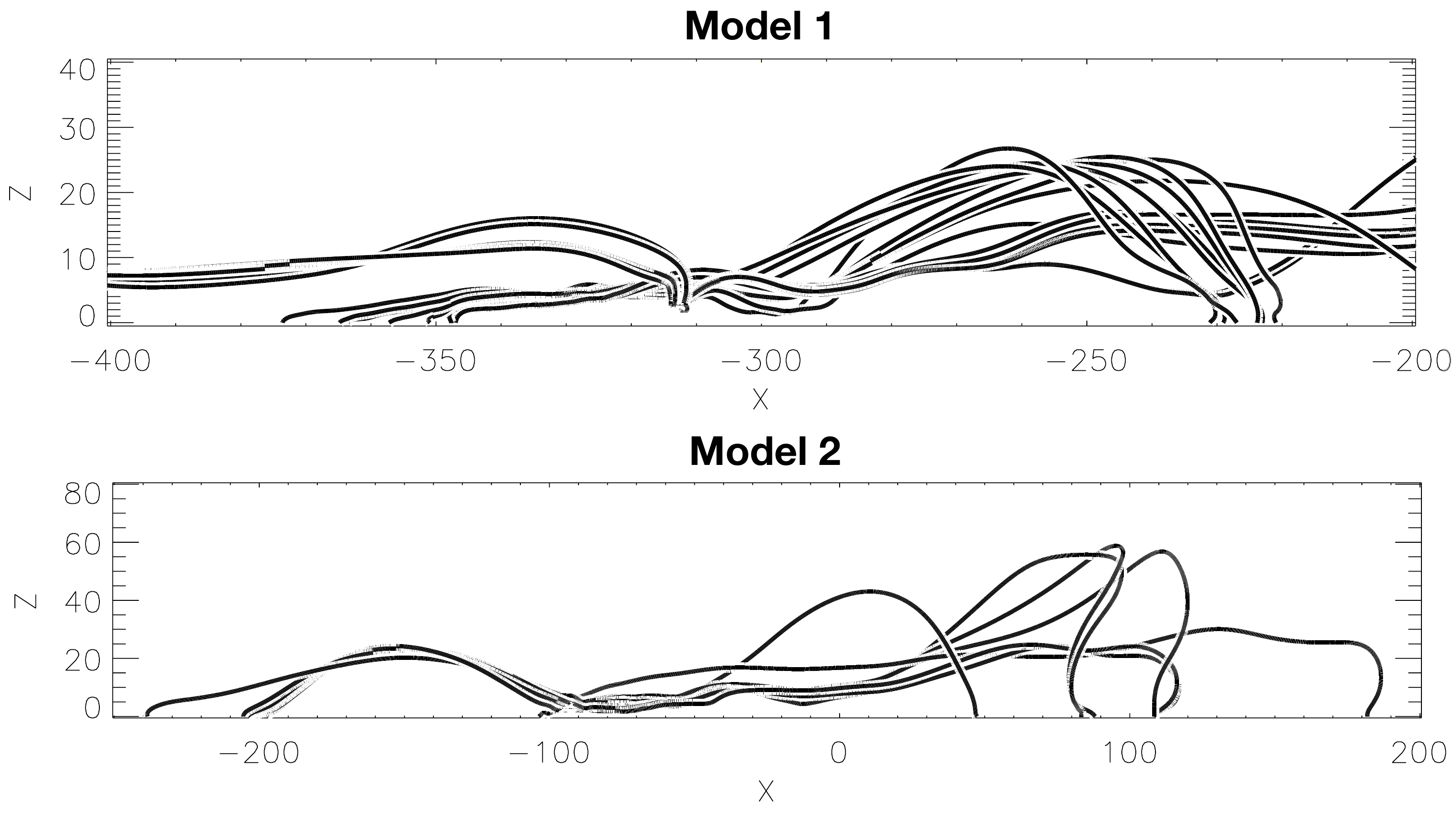}
\caption{ Sample magnetic field lines from models 1 and 2 plotted in the x-z plane to show the magnetic topology of the modelled flux ropes.} \label{fig7}
\end{figure*}

Figure~\ref{fig6} shows sample field lines taken from the best-fit magnetic models. The field lines in model 1 are relatively twisted and confined along the PIL of AR 11541 whereas, in the quiet sun region that extends towards AR 11538, the field lines appear less twisted. The field lines along the internal PIL of AR 11541 connect across the two main sites of flux cancellation. Conversely, in model 2 the field lines are more continuous, especially along the PIL of AR 11541 and at the western end that extends towards AR 11538. This is consistent with the evolution of the filament in 304~{{\AA}} where flows are seen along the eastern end and the filament at the western end grows in length. Sample field lines plotted in the x-z plane show that a bald patch topology present in both of the modelled flux ropes (see Figure~\ref{fig7}).

\subsection{Location of Modelled Dips}

\begin{figure*}[t]
\epsscale{1}
\plotone{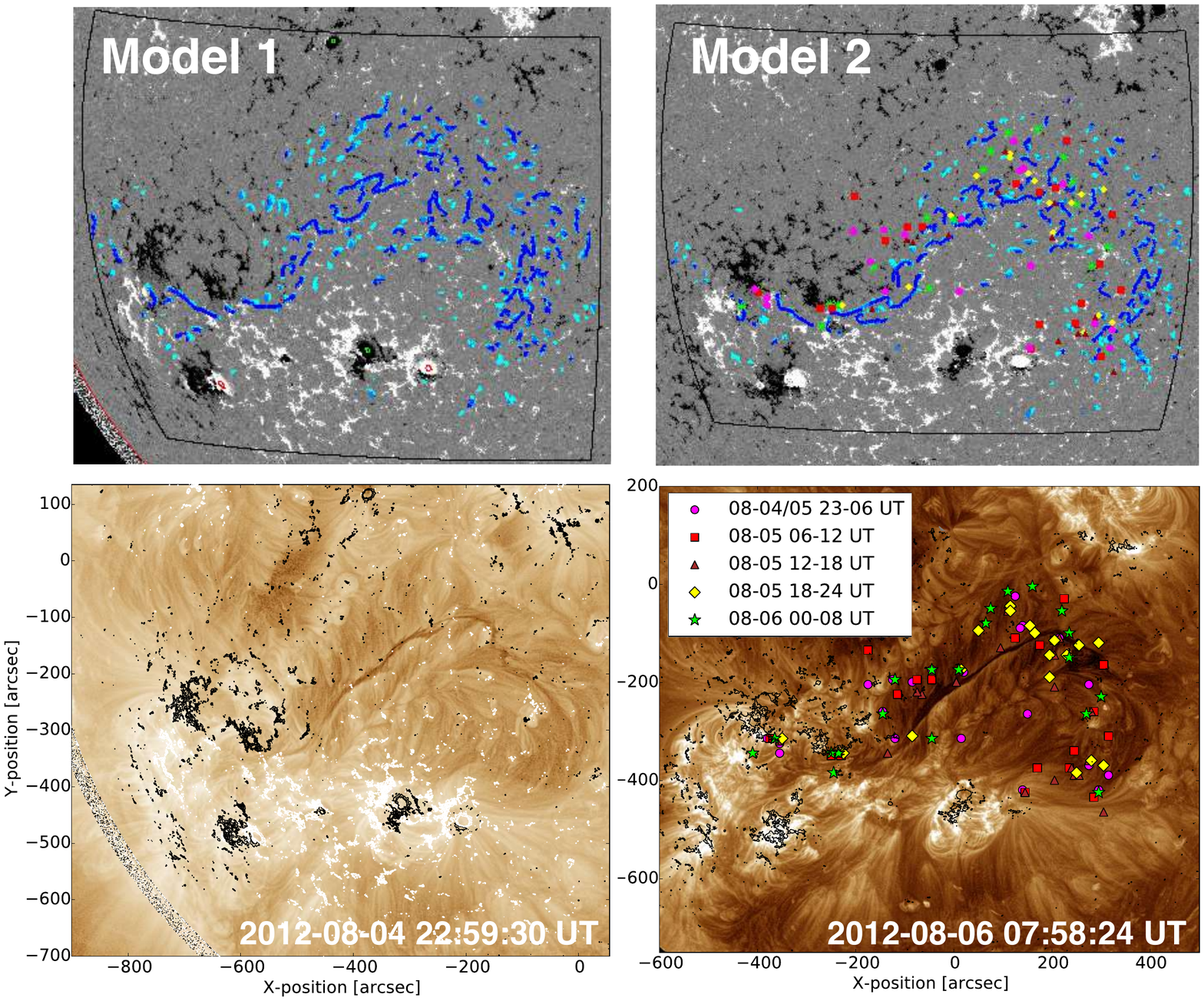}
\caption{The magnetic dips taken from NLFFF models 1 and 2 (2012 August 4 23:00~UT and 2012 August 6 08:00~UT) and the corresponding \textit{SDO}/AIA 193~{\AA} images of the filament. The magnetic dips are plotted at heights of 4 (light blue) and 6 (blue)~Mm. The magnetic dips and EUV images of the filament are overlaid on \textit{SDO}/HMI magnetograms at the times of the models with saturation levels of $\pm$~100~G. The location of the flux cancellation sites are plotted in the top right panel in white and the bottom right panel, where the different markers represent cancellation occurring during $\sim$6~hour time periods given in the legend.
}
\label{fig8}
\end{figure*}

One aim of constructing the NLFFF models is to investigate whether the location of magnetic dips is consistent with the observed flux cancellation sites and how these both evolve over the time period between the first model to the second. The location of the dips in the modelled magnetic field is shown in Figure~\ref{fig8}. Dips at a height of 4 and 6~Mm are plotted in light blue and blue, respectively. These dips are lower than the altitude of the lowest observed filament plasma, which is estimated from the observations to be 7~Mm. Magnetic dips at heights lower than the filament were chosen in order to investigate whether they correspond to the flux cancellation sites in the observations. These concave-up sections of magnetic field may be sites where filament plasma eventually accumulates.

The dips are observed along the PIL at the locations of the filament material and the surrounding flux rope with the dips being more broadly distributed in area in the quiet sun. There are fewer magnetic dips present in the cell volume in model 2 compared to model 1, however, the dips that remain are concentrated along the PIL underneath the filament plasma.
There is a change in the location of dips at the western end of the filament in the quiet sun, where the filament is observed to have accumulated more plasma, as seen in H$\alpha$ and 304~{\AA} (Figure~\ref{fig3} and \ref{fig4}) and flux cancellation has taken place. The dips along the filament spine appear to be fragmented due to line-of-sight effects and the fact that the dips are only plotted between the heights of 4 and 6~Mm.

\subsection{Current Distribution}

\begin{figure*}[t]
\epsscale{1}
\plotone{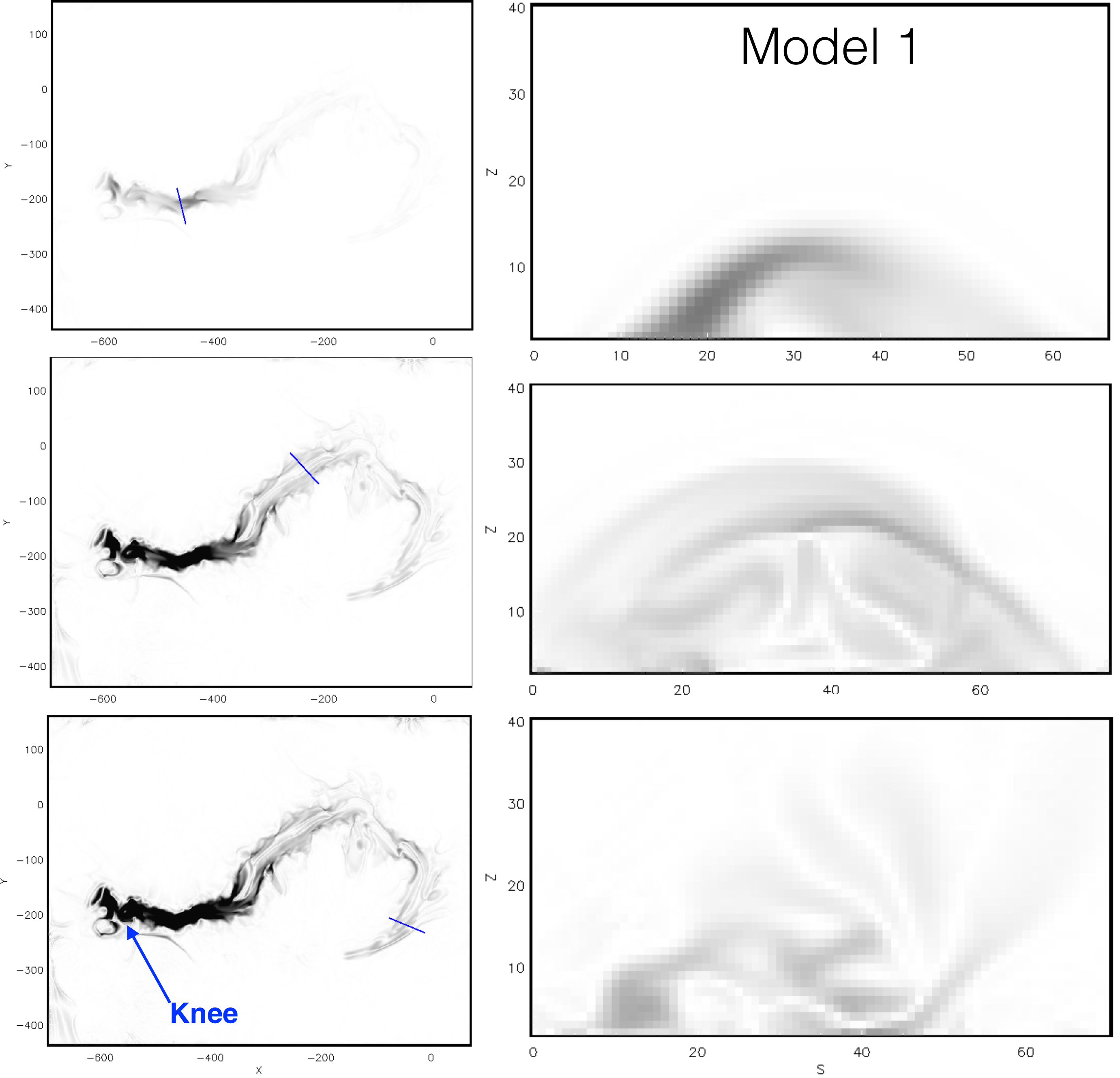}
\caption{Current density maps from NLFFF model 1 taken at 10,450~km above the photosphere. The blue lines in the left column represents the location of the cross sections given in the right column. Cross sections are taken in the three different sections of the filament as defined by the filament barbs. The coordinates are given in model coordinates and represent the number of cells where each cell is equivalent to 0.0015~R$_{\odot}$. 
The blue arrow shows the ``knee" of the filament, which is also present in the observations. } \label{fig9}
\end{figure*}

\begin{figure*}[t]
\epsscale{1}
\plotone{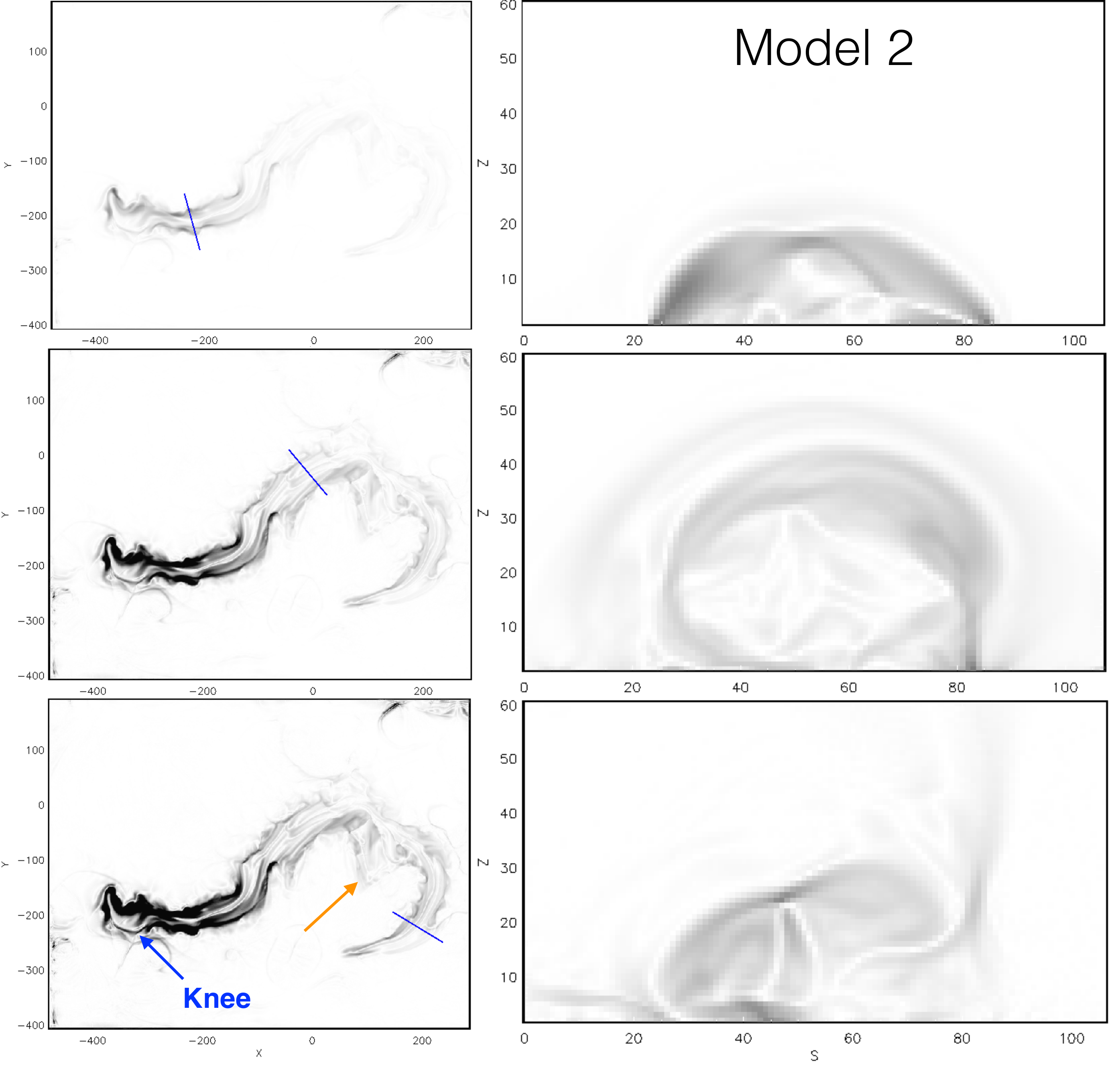}
\caption{Same as Figure~\ref{fig9} but for the second NLFFF model. 
The orange arrow in the bottom left panel shows an area of enhanced current corresponding to the locations of a large number of observed flux cancellation sites present in the quiet sun. } \label{fig10}
\end{figure*}

Figures~\ref{fig9} and \ref{fig10} show the current density maps of the two NLFFF models at different locations along each flux rope. In both models it is evident that the flux rope has a complex structure and in the western end of the rope the current is concentrated in the outer layers of the rope. The distribution of current of the eastern section of the flux rope, which is located in the AR, is very compact whereas the middle and western sections that are located in the quiet sun are very broad. The ``knee" in the eastern section of the filament, that is present in the observations, is also visible in the current distribution (arrows in Figure~\ref{fig9} and Figure~\ref{fig10}). In Figure~\ref{fig10}, which shows model 2, the current distribution is enhanced in the western end compared to the previous model. This is consistent with the growth in the filament and the location of a large number of flux cancellation sites in the quiet sun in the observations (orange arrow in Figure~\ref{fig10}).

\section{Summary \& Discussion} \label{sec:sum}

We present an analysis of an intermediate filament, from both an observational and a modelling perspective. The intermediate filament was located along a photospheric PIL that at the western end was rooted in the periphery of the positive polarity of AR 11538, extended through a quiet sun region, and ended in the negative polarity of AR 11541. Photospheric and atmospheric observations over a 33 hour time period were analyzed and supplemented by NLFFF models that were constructed at the start and end of this time period. The aim of the analysis is to investigate the magnetic field configuration that supports the filament plasma and probe the physical processes that underlie the formation and evolution of filaments. In particular, we ask whether the process known as flux cancellation plays an important role in the filament’s evolution.

The filament formed in the southern hemisphere and was determined to be of sinistral chirality from the orientation of its barbs as seen in H$\alpha$ observations. This is in agreement with the study of this filament by \citet{Joshi-2014} and is typical for southern hemisphere filaments in general. The magnetic field configuration of the filament was investigated through the construction of NLFFF models on 2012 August 4 at 23:00~UT and on 2012 August 6 at 08:00~UT. The models were created using the flux rope insertion method \citep{vB-2004}, which utilizes observations of a filament to guide the insertion of a weakly-twisted flux rope into a potential field extrapolation that is created using the LoS photospheric magnetic field as the boundary condition. The modelled field is then relaxed to remove Lorentz forces. The axial and poloidal flux of the flux rope can be varied so that the relaxed magnetic field lines are seen to match observed plasma emission structures. The magnetic field environment of the intermediate filament studied here (involving active region and quiet sun field) resulted in the modelled flux rope being composed of three sections, each with varying axial flux. These sections are referred to as the eastern, central and western sections and will be discussed more later. The two flux rope models, validated against observations, suggest that the filament plasma was supported by a weakly-twisted flux rope, which had right handed twist in line with its sinistral categorization.

Validating the NLFFF models for this intermediate filament was challenging. The extension of the filament, and its associated channel that is dark in EUV, into the quiet sun provides few plasma emission structures that can be used to select the modelled field configuration that best matches the observed coronal loops. To overcome this challenge we use EUV observations at times close to, but not the same as, the times of the NLFFF model. In particular we utilize times when the filament was activated approximately 12 hours before the first NLFFF model and 6 hours after the second NLFFF model. At these times fine threads of plasma become heated, revealing the global structure of the filament.

Given the flux rope magnetic field configuration we now discuss the position of magnetic dips in the NLFFF models and the sites of flux cancellation. Dips in the modelled field at a relatively low altitude of 4 and 6~Mm were investigated, with a focus on whether there is an overall change in the distribution of dips between the first and second NLFFF models. Overall, the dips in the modelled field are seen to be concentrated along the photospheric PIL in both models, where it is presumed that flux cancellation has taken place prior to the time period of this study. Comparing the first NLFFF model to the second model allows us to investigate the difference in the magnetic field over time and reveals that the number of dips in the cell volume decreases from the first to the second model. However, the dips that remain are concentrated along the PIL underneath the filament plasma.

There is a reduction in the number of field line dips in the region that corresponds to sites of flux cancellation away from the PIL, for example, in the quiet sun region on the filament's northern side. Flux cancellation in these locations presumably involves smaller-scale flux systems, that do not themselves build into the main filament magnetic structure and do not contribute axial field. It may be that in these locations dips are short-lived, or they may be created at different altitudes to the ones that we study. The results presented here emphasis the importance of ongoing flux cancellation at the same PIL to build up the non-potential structure that supports the filament plasma, which in this study forms a flux rope.


The flux cancellation sites that occur close to the polarity inversion line are concentrated at the eastern and western ends of the filament, in active region and decayed field regions respectively. The flux cancellation that occurred along the internal PIL of AR 11541, at the filament’s eastern end, was quantified by measuring the reduction in total positive magnetic flux during the time period beginning 2012 August 4 at 23:00~UT until 2012 August 6 at 08:00~UT. This time period coincides with the time when the filament was close to central meridian, meaning that our results are unaffected by HMI instrument sensitivity issues that cause peaks in the line of sight field for flux measured at a centre-to-limb angle of 60$^{\circ}$ \citep{Hoeksema-2014, Couvidat-2016}. The geosynchronous orbit of {\it SDO} also introduces a time-varying systematic error in the flux that manifests itself as superimposed sinusoidal oscillations with periods of 12 and 24~hours \citep{Hoeksema-2014}, compared to our time period of study of 33 hours. Overall, the total positive flux cancelled along the internal PIL of AR 11541 was calculated to be 3.2~$\times$~10$^{20}$~Mx. This is in fairly good agreement with the increase in flux in the eastern section of the modelled flux rope, which was found to be 4$\times$10$^{20}$~Mx. 

It was not possible to quantify the amount of flux cancelled in the western section of the flux rope due to the large quantity and small size of the fragments there. However, the growth of the western section of the filament does correspond to an area in the quiet sun where a large number of flux cancellation sites are observed (12 sites in $\sim$5000~Mm$^{2}$) and an increase in the size of the filament is observed. \citet{Li-2013} were able to measure the magnetic flux at the footpoints of fine-scale structures that are part of the western end of the flux rope at the time of the filament activation at around 12:56~UT on 2012 August 4. The footpoints of the fine-scale structures have values of magnetic flux ranging from 1.1~$\times$~10$^{19}$~Mx to 8.1~$\times$~10$^{19}$~Mx, and \citet{Li-2013} deduced that the flux rope contains at least 7.6~$\times$~10$^{20}$~Mx. The NLFFF model on this day suggests a lower value of total flux of 0.5$\times$10$^{20}$~Mx contained in the western end of the flux rope, which is an order of magnitude less than \citet{Li-2013}. This is due to the model containing three sections that are joined to create one flux rope. If we insert one flux rope into the model then the flux rope is no longer in equilibrium, which does not reflect the filament seen in the observations. The NLFFF model on August 6 contains a larger amount of flux in all sections of the flux rope. This suggests that the cancellation of flux builds flux into the flux rope. These observations are consistent with the previous studies of \citet{Liu-2012, Zhang-2014, Yardley-2016}, where dense filamentary threads build into a filament through the process of reconnection that is associated with flux cancellation.

Whilst the evolution of the axial field of the modelled flux rope bears close correspondence with the observed flux cancellation, the poloidal flux does not change between the models. This is driven by the observations that show no support for the structure becoming more twisted with time. Although flux cancellation builds axial flux into the rope, it doesn't necessarily increase the poloidal flux. It depends on the details of how the flux cancellation proceeds. For details see \citet{Green-2011}.

The height range of plasma in the central section of the filament was measured to be between 7 and 46~Mm using observations from STEREO-B, which was close to being in quadrature at the time of this study. This height range is higher than the dips in the field that we investigate using the NLFFF models. However, the close spatial correspondence of the dipped field, flux cancellation and the filament plasma locations as seen from above is still consistent with the plasma being supported in dips in the field that were created as a result of flux cancellation. The NLFFF models allow us to probe in more detail the overall height and extent of the flux rope at the start and end of our time period of interest. Over the 33 hours from the time of the first model to the second, all three sections of the modelled rope show an increase in altitude with the rise being most pronounced in the central and western sections. This is consistent with the observations that show the eruption of the western end of the filament occurs $\sim$6 hours after the second NLFFF model and causes part of the filament and supporting structure to rise and become destabilized.

The modelling results in our work are consistent with the observations that show that the eventual eruption of the filament initiates in the western (quiet sun) section. In the western section the magnetic structure is evolving due to flux cancellation and, as indicated from the modelling, the axial flux doubles in size accompanied by a rise of the structure. Both the flux cancellation, which will add axial flux to the rope, and the rise of the structure are consistent with the eventual eruption of the filament. The lack of helical deformation of the filament in the observations in the lead up to eruption along with the low values of poloidal flux in the NLFFF models leads us to believe that the eruption is not due to the kink instability \citep{Torok-2005}. The onset of eruption is therefore most likely caused by the slow rise of the filament to a height at which the gradient of the magnetic field overlying the filament decays rapidly enough with height for the torus instability to occur \citep{Kliem-2006, Demoulin-2010, Kliem-2013}.

In summary, observations of an intermediate filament were analyzed and compared to the NLFFF models constructed using the flux rope insertion method. The location and evolution of plasma distribution, sites of flux cancellation and magnetic dips in the NLFFF models were found to be in agreement. In addition, the total flux cancelled as calculated from the HMI observations was comparable to the change in axial flux between the NLFFF models. Finally, there was an increase in the free magnetic energy 0.2$\times$10$^{31}$~ergs between the NLFFF models. Therefore, these results are consistent with flux cancellation and associated reconnection lifting plasma into a forming or growing flux rope, lengthening the pre-existing filament and producing flows along the filament that originate from the flux cancellation sites.

\acknowledgements

The authors are grateful to the {\it SDO}/HMI and AIA consortia for the data. This research makes use of SunPy \citep{SunPy-2015}, an open-source and free community-developed solar data analysis package written in Python. S.L.Y. is grateful to STFC for support via PhD studentship and the Consolidated Grant SMC1/YST025. L.M.G. acknowledges support through a Royal Society University Research Fellowship. L.v.D.G. is partially funded under STFC consolidated grant number ST/N000722/1. L.v.D.G. also acknowledges the Hungarian Research grant OTKA K-109276. L.M.G. and L.v.D.G. acknowledge funding under Leverhulme Trust Research Project Grant 2014-051. D.H.M. would like to thank both the STFC and Leverhulme trust for financial support. D.M.L. 
D.M.L. acknowledges support from the European Commission's H2020 Programme under the following Grant Agreements: GREST (no.~653982) and Pre-EST (no.~739500) as well as support from the Leverhulme Trust for an Early-Career Fellowship (ECF-2014-792) and is grateful to the Science Technology and Facilities Council for the award of an Ernest Rutherford Fellowship (ST/R003246/1).

\end{document}